%% file: ICSE-FORGE-TypeEvalPy-LLMs.tex
\documentclass[sigconf,nonacm]{acmart}

\AtBeginDocument{%
  }

\acmConference[FORGE]{}{}{}
\acmISBN{978-1-4503-XXXX-X/18/06}

\input{packages.tex}

\begin{document}




\title[The Emergence of Large Language Models in Static Analysis: A First Look through Micro-Benchmarks]{The Emergence of Large Language Models in Static Analysis: A First Look through Micro-Benchmarks}


\author{Ashwin Prasad Shivarpatna Venkatesh\textsuperscript{§}, 
	Samkutty Sabu\textsuperscript{¶},
	Amir M. Mir\textsuperscript{‡},
	Sofia Reis\textsuperscript{†},
	Eric Bodden\textsuperscript{**}}
\affiliation{%
	\textsuperscript{§}\textit{ashwin.prasad@upb.de}, Heinz Nixdorf Institut, Paderborn University, Paderborn, Germany\\
	\textsuperscript{¶}\textit{samkutty@mail.uni-paderborn.de}, Paderborn University, Paderborn, Germany\\
	\textsuperscript{‡}\textit{s.a.m.mir@tudelft.nl}, Delft University of Technology, Delft, The Netherlands\\
	\textsuperscript{†}\textit{sofia.o.reis@tecnico.ulisboa.pt}, IST, University of Lisbon \& INESC-ID, Lisbon, Portugal\\
	\textsuperscript{**}\textit{eric.bodden@upb.de}, Heinz Nixdorf Institut \& Fraunhofer IEM, Paderborn University, Paderborn, Germany\\
	\country{}
}

\renewcommand{\shortauthors}{Venkatesh et al.}

\newcommand{\headergen}{\textsc{HeaderGen}\xspace}
\newcommand{\typeevalpy}{\textsc{TypeEvalPy}\xspace}
\newcommand{\pycg}{\textsc{PyCG}\xspace}

\hyphenation{Header-Gen}
\hyphenation{Type-EvalPy}
\hyphenation{Py-CG}
\begin{abstract}
The application of Large Language Models (LLMs) in software engineering, particularly in static analysis tasks, represents a paradigm shift in the field.
In this paper, we investigate the role that current LLMs can play in improving callgraph analysis and type inference for Python programs. 
Using the PyCG, HeaderGen, and TypeEvalPy micro-benchmarks, we evaluate 26 LLMs, including OpenAI's GPT series and open-source models such as LLaMA.
Our study reveals that LLMs show promising results in type inference, demonstrating higher accuracy than traditional methods, yet they exhibit limitations in callgraph analysis. 
This contrast emphasizes the need for specialized fine-tuning of LLMs to better suit specific static analysis tasks.
Our findings provide a foundation for further research towards integrating LLMs for static analysis tasks.
\end{abstract}

\maketitle

\input{introduction.tex}
\input{motivating_example.tex}

\input{rw.tex}

\input{RQ.tex}

\input{methodology.tex}
\input{results.tex}

\input{discussions.tex}

\input{future_work_conclusion.tex}

\begin{acks}
	Funding for this study was provided by the Ministry of Culture and Science of the State of North Rhine-Westphalia under the SAIL project with the grand no NW21-059D.
\end{acks}

\bibliographystyle{ACM-Reference-Format}
\bibliography{references}

\end{document}

%% file: packages.tex
\usepackage{xspace}
\usepackage{booktabs} 
\usepackage{siunitx} 
\usepackage{tabularx} 
\usepackage{xspace}
\usepackage{multirow}
\usepackage{listings}
\usepackage{enumitem}
\usepackage{tikz}
\usepackage{colortbl}

\usepackage[colorinlistoftodos,prependcaption,textsize=tiny]{todonotes}

\definecolor{codegreen}{rgb}{0,0.6,0}
\definecolor{codegray}{rgb}{0.5,0.5,0.5}
\definecolor{codepurple}{rgb}{0.58,0,0.82}
\definecolor{backcolour}{rgb}{0.95,0.95,0.92}

\lstdefinestyle{mystyle}{
	commentstyle=\color{codegreen},
	keywordstyle=\color{magenta},
	numberstyle=\tiny\color{codegray},
	stringstyle=\color{codepurple},
	basicstyle=\small,
	breakatwhitespace=false,         
	breaklines=true,                 
	captionpos=b,                    
	keepspaces=true,                 
	numbers=left,                    
	numbersep=5pt,                  
	showspaces=false,                
	showstringspaces=false,
	showtabs=false,                  
	tabsize=4,
}

%% file: introduction.tex
\section{Introduction}

In the dynamic field of Software Engineering (SE), the incorporation of advanced computational models, especially Large Language Models (LLMs), marks a significant shift in the software development processes \cite{houLargeLanguageModels2023, fanLargeLanguageModels2023, zhangSurveyLargeLanguage2023, zhengUnderstandingLargeLanguage2023}.
Static analysis (SA), an integral component of SE, involves examining source code without executing it, to identify potential errors, code quality issues, and security vulnerabilities. 
The emergence of LLMs, such as BERT~\cite{devlinBERTPretrainingDeep2019a}, T5~\cite{raffelExploringLimitsTransfer2023}, and GPT~\cite{radfordLanguageModelsAre}, has transformed several diverse SE tasks, including the SA tasks~\cite{zhangSurveyLargeLanguage2023}.
Recent works have shown how different SA tasks can benefit from LLMs, such as false-positives pruning~\cite{10.1145/3611643.3613078}, improved program behavior summarization~\cite{liHitchhikerGuideProgram2023}, type annotation ~\cite{seidel2023learning}, and general enhancements in precision and scalability of SA tasks~\cite{liHitchhikerGuideProgram2023}, both fundamental issues of SA.

This study here situates itself at the intersection of SA and LLMs, specifically focusing on the effectiveness of LLMs in SA within SE.
It aims to evaluate the accuracy of LLMs in performing specific SA tasks: callgraph analysis and type inference, specifically in Python programs.
\textit{Callgraph analysis} helps in understanding the relationships and interactions between different components of a program, while \textit{type inference} aids in identifying potential type errors and improving code reliability.
To assess the performance of LLMs in these areas, we use the \pycg~\cite{salisPyCGPracticalCall2021c} and 
\headergen~\cite{venkateshEnhancingComprehensionNavigation2023a} micro-benchmarks for callgraph analysis, and \typeevalpy~\cite{venkateshTypeEvalPyMicrobenchmarkingFramework2023} for type inference. 

The use of micro-benchmarks in evaluating the performance of LLMs in our study is grounded in several key considerations.
Firstly, micro-benchmarks are designed to target specific aspects of the features under test and various characteristics of the programming language involved. 
This helps highlight the models' strengths and weaknesses, allowing for a more nuanced understanding of their capabilities in SA tasks.
Additionally, their development involves rigorous manual inspection and adherence to scientific methods, ensuring reliability and accuracy in evaluation. 
Conversely, obtaining large-scale, real-world data that can serve as ground truth is often a challenging endeavor. 
Where such data is available, it is susceptible to human errors, which can skew the results. 

By testing a range of $26$ different LLMs, our study provides a comprehensive analysis of their capabilities in the context of SA.
Furthermore, the evaluation enables one to make direct comparisons with the existing capabilities of traditional approaches in SA.
The insights from this study are intended to offer a preliminary understanding of the role LLMs can play in SA, and potentially guide future research and practical applications in the AI4SE and SE4AI fields.

The structure of the paper is as follows: Section~\ref{sec:background} provides a motivating example to introduce the concepts of SA.
In Section~\ref{sec:relatedwork} we discuss the related work.
The research questions are outlined in Section~\ref{sec:rq}, while Section~\ref{sec:methodology} describes our methodology.
Results are presented in Section~\ref{sec:results} and subsequently discussed in Section~\ref{sec:discussion}. 
Section~\ref{sec:ttv} addresses the threats to validity.
Finally, the paper is concluded by outlining future research directions in section~\ref{sec:conclusion}.

\textbf{Availability.}
\typeevalpy is published on GitHub as open-source software:
\url{https://github.com/secure-software-engineering/TypeEvalPy}

%% file: motivating_example.tex
\section{Background}
\label{sec:background}

In the following code, the \texttt{create\_str} function returns a string, the variable \texttt{func\_ref} is assigned with function references at lines 4 and 8 and \texttt{x} is assigned the value \texttt{result + 1} at lines 6 and 10. 

\begin{lstlisting}[language=Python]
def create_str(x):
	return x.upper()

func_ref = create_str
result = func_ref("Hello!")
x = result + 1 # Type mismatch!

func_ref = len
result = func_ref("Hello!")
x = result + 1 # Works
\end{lstlisting}

\textbf{Type Inference.}
A static analyzer with type inference capabilities can resolve that the variable \texttt{result} at line 5 is a string, while the variable \texttt{result} at line 9 is an integer.
Using this, the static analyzer can raise a type error at line 6 even before executing it.

\textbf{Callgraph.}
The complete callgraph for the snippet is as follows:

\textit{main $\rightarrow$ create\_str() $\rightarrow$ upper()}

\textit{main $\rightarrow$ len()}

A flow-sensitive analysis can further resolve exactly where these calls are made.
For instance, it can resolve that at line 5 the variable \texttt{func\_ref} points to the function \texttt{create\_str} while  at line 9 \texttt{func\_ref} points to the function \texttt{len}.

%% file: rw.tex
\section{Related Work}
\label{sec:relatedwork}

Ma et al.~\cite{scopeofGPT} and Sun et al.~\cite{sun2023automatic}
explore the capabilities of LLMs when performing different program
analysis tasks such as control-flow graph construction, callgraph
analysis, and code summarization. They conclude that while LLMs
can comprehend basic code syntax, they are somewhat limited in
performing more sophisticated analyses, such as pointer analysis
and code behavior summarization. In contrast, LLift, an LLM-based approach, showed successful results for different programming analysis tasks, including program behavior summarization~\cite{liHitchhikerGuideProgram2023} and how LLMs can be successfully integrated into an SA pipeline. Researchers conjecture that the reasons behind the difference in the results were benchmark selection, prompt designs, and model versions. Li et al.~\cite{10.1145/3611643.3613078} present a solution to prune SA false positives by asking carefully constructed questions about function-level behaviors or function summaries.
Seidel et al.~\cite{seidel2023learning} propose CodeTIDAL5, a Transformer-based model trained to predict type annotations in TypeScript. In this study, we explore how different LLMs perform on callgraph analysis and type inference for Python programs.

%% file: RQ.tex
\section{Research Questions}
\label{sec:rq}

We focus on the following research questions to evaluate the effectiveness of LLMs using micro-benchmarks in static analysis tasks: 

\begin{itemize}	
	\item[\textit{\textbf{RQ1:}}] \textit{What is the accuracy of LLMs in performing callgraph analysis against micro-benchmarks?}
	\item[\textit{\textbf{RQ2:}}] \textit{What is the accuracy of LLMs in performing type inference against micro-benchmarks?}
\end{itemize}

%% file: methodology.tex
\section{Methodology}
\label{sec:methodology}
We next describe the experimental setup, the model selection criteria, prompt design, and metrics used to investigate these RQs.

\textbf{Micro-benchmarks.}
To answer RQ1, we choose two benchmarks designed to evaluate callgraph analysis performance, \pycg~\cite{salisPyCGPracticalCall2021c} and \headergen~\cite{venkateshEnhancingComprehensionNavigation2023a}.
\pycg is the first callgraph construction algorithm that uses a context-\textit{insensitive} and flow-\textit{insensitive} SA as its backend.
\pycg includes a micro-benchmark containing 112 unique python programs targeting various Python features organized into 16 categories.
\headergen is a tool that uses SA to enhance comprehension in computational notebooks.
\headergen improves \pycg's static analyzer with flow-sensitivity and type inference.
\headergen includes a micro-benchmark with 121 code snippets with flow-sensitive call sites as ground truth.
Note that for this study we have extended \pycg's micro-benchmark with additional snippets from the \headergen micro-benchmark.

To answer RQ2, we choose the micro-benchmark from \typeevalpy~\cite{venkateshTypeEvalPyMicrobenchmarkingFramework2023}, a general framework for evaluating type inference tools in Python.
\typeevalpy contains a micro-benchmark with 154 code snippets and 845 type annotations as ground truth.

\textbf{Model Selection.}
In this study, we evaluate several state-of-the-art LLMs.
First, we include two closed-source LLMs, GPT 3.5 Turbo and GPT 4 from OpenAI as it is the leading general-purpose LLM.
Furthermore, we include ten popular open-source models based on the download count on the Huggingface \cite{HuggingFaceAI} platform.
This includes llama2, mistral, dolphin-mistral, codellama, codebooga, tinyllama, vicuna, wizardcoder, and orca.
We include several variations of these models such as the number of parameters (7b, 13b, etc.,).
Overall, we evaluate 24 open-source models and two closed-source models, totaling 26 LLMs.

Furthermore, we create a fine-tuned version of GPT-3.5 Turbo, refined with a training dataset.
The dataset created for fine-tuning GPT-3.5 Turbo comprises 15 program categories.
It serves as the representative collection of the \pycg, \headergen, and \typeevalpy micro-benchmarks, emphasizing key Python features such as functions, classes, decorators, and exceptions.
This approach seeks to enhance the model's adaptability, equipping it to effectively handle a diverse range of challenges.

\textbf{Prompt Design.}
To optimize prompt design, we adopted an iterative and experimental approach \cite{chenUnleashingPotentialPrompt2023}.
Initial efforts focused on enhancing the prompt by incorporating detailed task descriptions and specifying the expected response format.
Notably, we used a one-shot prompting technique, embedding an example question and answer within the prompt.
Despite these refinements, we encountered challenges with the LLM's ability to produce \emph{structured} outputs.
Our experiments revealed that even with explicit instructions to generate outputs in JSON format, models struggled to deliver results that could be reliably parsed.
To address this, we explored a question-answer based method, querying the model and then translating its natural-language responses back into a structured JSON format. This offers a more flexible solution to the challenges of generating structured data outputs.


\textbf{Evaluation Metrics.}
To assess both flow-insensitive callgraph construction and flow-sensitive call-site extraction, in this study, we measured completeness, soundness, and exact matches. 
Completeness is the absence of false positives in the callgraph, ensuring that no call edges were included if they did not exist.
Soundness, conversely, focuses on the inclusion of every call edge, thereby avoiding any false negatives. 
Exact matches is measured as the number of function calls that exactly match the ground truth.
This evaluation approach mirrors the methodologies used in previous studies, specifically in \pycg~\cite{salisPyCGPracticalCall2021c} and \headergen~\cite{venkateshEnhancingComprehensionNavigation2023a}. 
Furthermore, aligning with the literature \cite{Typilus, mirType4PyPracticalDeep2022c, HiTyper, venkateshTypeEvalPyMicrobenchmarkingFramework2023}, for type-inference evaluation we use exact matches as the metric.
Additionally, the total runtime of these tools for analyzing the respective micro-benchmark is also included by computing the mean over three runs.

\textbf{Implementation Details.}
In the implementation of our experiments with LLMs, we employed Ollama \cite{Ollama}, an open-source platform that simplifies running LLMs by providing an efficient HTTP server for lifecycle management.
This served as our backend infrastructure. 
In addition, to create a pipeline for efficient prompting and response handling, we used LangChain \cite{LangchainaiLangchainBuilding}, a framework designed for building applications that interact with LLMs.
Additionally, to implement the type-inference experiments, we extended the TypeEvalPy framework~\cite{venkateshTypeEvalPyMicrobenchmarkingFramework2023}, due to its flexibility in adding support for new tools.

%% file: results.tex
\section{Results}
\label{sec:results}
We next present the findings of our study, addressing the research questions and highlighting key results.

\begin{table*}[h]
	\vspace{-.2cm}
	\renewcommand{\arraystretch}{1.1}
	\setlength{\tabcolsep}{3.5pt} 
	\caption{Comparative analysis across LLMs for callgraph analysis on \pycg and \headergen micro-benchmarks}
	\label{tab:cg_cs}

	\vspace{-.3cm}	
	\begin{tabular}{lrrrrllrrrr}
		\cmidrule(lr){1-5} \cmidrule(lr){7-11}
		\multicolumn{5}{c}{\textbf{\pycg Benchmark --- Flow-insensitive Callgraphs }}                                                                                                                        &  & \multicolumn{5}{c}{\textbf{\headergen Benchmark --- Flow-sensitive Callgraphs}}                                                                                                                                \\
		\cmidrule(lr){1-5} \cmidrule(lr){7-11}

\multicolumn{1}{c}{\textbf{Model}}    & \multicolumn{1}{c}{\textbf{Complete}} & \multicolumn{1}{c}{\textbf{Sound}}   & \multicolumn{1}{c}{\textbf{E.M}}     & \multicolumn{1}{c}{\textbf{Time}}     &  & \multicolumn{1}{c}{\textbf{Model}}         & \multicolumn{1}{c}{\textbf{Complete}} & \multicolumn{1}{c}{\textbf{Sound}}   & \multicolumn{1}{c}{\textbf{E.M}}     & \multicolumn{1}{c}{\textbf{Time}}      \\
\cellcolor[HTML]{EFEFEF}\textbf{PyCG} & \cellcolor[HTML]{EFEFEF}\textbf{113}  & \cellcolor[HTML]{EFEFEF}\textbf{105} & \cellcolor[HTML]{EFEFEF}\textbf{250} & \cellcolor[HTML]{EFEFEF}\textbf{0.41} &  & \cellcolor[HTML]{EFEFEF}\textbf{HeaderGen} & \cellcolor[HTML]{EFEFEF}\textbf{111}  & \cellcolor[HTML]{EFEFEF}\textbf{113} & \cellcolor[HTML]{EFEFEF}\textbf{327} & \cellcolor[HTML]{EFEFEF}\textbf{5.26} \\
ft:gpt-3.5-turbo                      & 70                                    & 75                                   & 207                                  & 77.96                                 & \multicolumn{1}{l}{} & ft:gpt-3.5-turbo                           & 47                                    & 48                                   & 149                                  & 79.98                                  \\
gpt-4                                 & 59                                    & 54                                   & 180                                  & 264.96                                &                      & gpt-4                                      & 27                                    & 17                                   & 70                                   & 248.37                                 \\
codebooga                             & 22                                    & 44                                   & 140                                  & 462.00                                &                      & gpt-3.5-turbo                              & 17                                    & 16                                   & 53                                   & 160.14                                 \\
phind-codellama:34b-v2                & 77                                    & 21                                   & 70                                   & 696.01                                &                      & phind-codellama:34b-v2                     & 13                                    & 12                                   & 42                                   & 475.08                                 \\
wizardcoder:7b-python                 & 21                                    & 18                                   & 60                                   & 157.11                                &                      & vicuna:13b                                 & 14                                    & 12                                   & 27                                   & 184.76                                 \\
wizardcoder:34b-python                & 76                                    & 13                                   & 45                                   & 847.04                                &                      & wizardcoder:34b-python                     & 8                                     & 8                                    & 17                                   & 360.71                                 \\
\textbf{codellama:34b-instruct}                & 2                                     & 7                                    & 40                                   & \textbf{1644.51}                      &                      & wizardcoder:13b-python                     & 12                                    & 9                                    & 15                                   & 195.17                                 \\
gpt-3.5-turbo                         & 14                                    & 20                                   & 40                                   & 124.58                                &                      & vicuna:33b                                 & 8                                     & 7                                    & 14                                   & 365.01                                 \\
orca2:13b                             & 14                                    & 20                                   & 39                                   & 386.47                                &                      & vicuna:7b                                  & 7                                     & 6                                    & 13                                   & 130.42                                 \\
codellama:13b-instruct                & 1                                     & 11                                   & 35                                   & 280.62                                &                      & llama2:7b                                  & 10                                    & 8                                    & 12                                   & 128.40                                 \\
wizardcoder:13b-python                & 6                                     & 9                                    & 29                                   & 232.74                                &                      & codebooga                                  & 6                                     & 8                                    & 11                                   & 357.18                                 \\
mistral:instruct                      & 12                                    & 7                                    & 28                                   & 188.85                                &                      & mistral:instruct                           & 7                                     & 7                                    & 10                                   & 191.66                                 \\
mistral:v0.2                          & 12                                    & 6                                    & 28                                   & 185.95                                &                      & tinyllama                                  & 10                                    & 6                                    & 10                                   & 355.92                                 \\
dolphin-mistral                       & 15                                    & 9                                    & 21                                   & 158.11                                &                      & codellama:34b-python                       & 15                                    & 10                                   & 9                                    & 270.90                                 \\
codellama:7b-instruct                 & 1                                     & 6                                    & 16                                   & 276.94                                &                      & dolphin-mistral                            & 11                                    & 11                                   & 9                                    & 129.36                                 \\
tinyllama                             & 28                                    & 8                                    & 13                                   & 889.33                                &                      & mistral:v0.2                               & 7                                     & 7                                    & 7                                    & 196.42                                 \\
orca2:7b                              & 106                                   & 7                                    & 10                                   & 336.35                                &                      & phind-codellama:34b-python                 & 11                                    & 8                                    & 7                                    & 319.54                                 \\
\textbf{vicuna:13b}                            & 3                                     & 9                                    & 9                                    & \textbf{2383.42}                      &                      & wizardcoder:7b-python                      & 8                                     & 8                                    & 7                                    & 135.16                                 \\
vicuna:7b                             & 1                                     & 8                                    & 8                                    & 147.05                                &                      & codellama:13b-python                       & 12                                    & 9                                    & 6                                    & 181.04                                 \\
vicuna:33b                            & 0                                     & 6                                    & 6                                    & 478.70                                &                      & codellama:7b-python                        & 11                                    & 9                                    & 5                                    & 300.54                                 \\
\textbf{llama2:70b}                            & 0                                     & 6                                    & 1                                    & \textbf{1398.26}                      &                      & orca2:13b                                  & 6                                     & 6                                    & 3                                    & 369.16                                 \\
codellama:13b-python                  & 121                                   & 0                                    & 0                                    & 142.09                                &                      & codellama:7b-instruct                      & 7                                     & 6                                    & 2                                    & 148.13                                 \\
codellama:34b-python                  & 121                                   & 0                                    & 0                                    & 269.77                                &                      & \textbf{llama2:70b}                                 & 6                                     & 6                                    & 1                                    & \textbf{944.54}                        \\
codellama:7b-python                   & 121                                   & 0                                    & 0                                    & 92.92                                 &                      & codellama:13b-instruct                     & 6                                     & 6                                    & 0                                    & 188.70                                 \\
llama2:13b                            & 14                                    & 6                                    & 0                                    & 587.44                                &                      & codellama:34b-instruct                     & 121                                   & 6                                    & 0                                    & 347.79                                 \\
\textbf{llama2:7b}                             & 93                                    & 0                                    & 0                                    & \textbf{1825.59}                      &                      & llama2:13b                                 & 6                                     & 6                                    & 0                                    & 426.48                                 \\
phind-codellama:34b-python            & 121                                   & 0                                    & 0                                    & 267.63                                &                      & orca2:7b                                   & 6                                     & 6                                    & 0                                    & 224.57                                

\\

		\cmidrule(lr){1-5} \cmidrule(lr){7-11}
	\end{tabular}
\vspace{-.25cm}
\end{table*}

\subsection{RQ1: Accuracy of Callgraph Analysis}

Table \ref{tab:cg_cs} presents the outcomes of our experiments using LLMs on the flow-insensitive callgraph analysis evaluation micro-benchmark of \pycg, and the flow-sensitive callgraph analysis evaluation micro-benchmark of \headergen.

\textbf{Flow-insensitive Callgraph analysis.}
The static analysis algorithm \pycg demonstrated superior performance over LLMs in terms of completeness, soundness, exact matches, and processing time.
Specifically, in a set of 121 test cases in the benchmark, \pycg achieved 93.3\% completeness and 86.7\% soundness, significantly outperforming the closest LLM, ft:gpt-3.5-turbo, which only achieved 57.8\% completeness and 61.9\% soundness.
Furthermore, \pycg obtained 250 exact matches (out of 284), which is 43 more exact matches than ft:gpt-3.5-turbo.
This performance difference is further emphasized in running times, where \pycg processed flow-insensitive callgraphs 190 times faster than ft:gpt-3.5-turbo.
Among the LLMs, the best-performing one without fine-tuning is gpt-4; however, the fine-tuned gpt-3.5-turbo model surpasses the vanilla gpt-4, indicating the potential benefits of fine-tuning LLMs for specific applications.
Yet, other open-source models lagged significantly in performance.
Notably, due to their failure to produce structured outputs in line with our prompts, some LLMs like codellama:34b-instruct, vicuna:13b, llama2:70b, and llama2:7b experienced lengthy running times. Despite clear instructions regarding the output format and the instruction to avoid explanatory content, they sometimes continued to generate irrelevant content and consequently reached the timeout.

\textbf{Flow-sensitive Callgraph analysis.}
Here, \headergen demonstrated superior performance over LLMs across all evaluated metrics.
In particular, \headergen achieved 91.7\% completeness and 93.3\% soundness, which is more than double the performance of its closest LLM competitor, ft:gpt-3.5-turbo, which managed only 38.8\% completeness and 39.6\% soundness.
In terms of exact matches, \headergen identified 327 out of 355 call sites, surpassing the best-performing LLM by 178 matches. Moreover, \headergen's runtime is 15 times shorter than the fastest LLM in analyzing the entire benchmark. 
Note that LLMs fared considerably poorer in flow-sensitive analysis compared to flow-insensitive analysis, likely due to the increased complexity and the requirement for precise flow-sensitive pointer information, which may pose challenges to LLMs. 
And this although in the prompt we did provide specific instructions to ensure the LLMs' awareness of the flow-sensitive aspects.

\input{sub_type_inference.tex}

%% file: sub_type_inference.tex
\subsection{RQ2: Accuracy of Type Inference}
\label{subsec:type-inf-res}
Table ~\ref{tab:type_inference} shows the performance of LLMs, HeaderGen, and HiTyper considering the exact-match performance. In general, LLMs significantly here outperform the current state-of-the-art approaches for type inference, namely, HeaderGen and HiTyper models. Specifically, OpenAI's GPT-4 is the best-performing model, correctly inferring 775 of 845 type annotations in the micro-benchmark. This is expected, as GPT-4 is one of the most powerful LLMs in the wild, though it can be slow and expensive to run. It is also interesting to see that the fine-tuned version of GPT 3.5 Turbo is the second best-performing model with 730 correctly inferred type annotations and an inference speed 4 times faster than that of GPT 4. Considering open-source LLMs, with 699 correctly inferred annotations CodeLlama (13B-instruct) has comparable performance to GPT-4 and the fine-tuned GPT 3.5. LLMs specialized in code-related tasks like CodeLLaMA outperform general-purpose LLMs such as vanilla LLaMA. Another observation is that TinyLlama, a 1.1B parameter model, performs poorly: it only infers 26 annotations correctly. It seems that models smaller than seven billion parameters, like TinyLlama, are insufficiently capable of the type inference task.

\begin{table}[H]
	\caption{Exact match comparison of LLMs in type inference}
	\label{tab:type_inference}
		\setlength{\tabcolsep}{3.0pt} 
	\renewcommand{\arraystretch}{1.1}
	
	\begin{tabular}{lrrrrr}
		\multicolumn{6}{r}{\Small \textbf{FRT:} Function return type, \textbf{FPT:} Function parameter type, \textbf{LVT:} Local variable type}   \\
		\toprule
		\multicolumn{1}{c}{\textbf{Model}} & \multicolumn{1}{c}{\textbf{FRT}} & \multicolumn{1}{c}{\textbf{FPT}} & \multicolumn{1}{c}{\textbf{LVT}} & \multicolumn{1}{c}{\textbf{Total}} & \multicolumn{1}{c}{\textbf{Time (s)}}\\
		\midrule
gpt-4                             & 225                             & 85                              & 465                             & 775                                & 454.54                   \\
ft:gpt-3.5-turbo                  & 209                             & 85                              & 436                             & 730                                & 110.45                        \\
codellama:13b-instruct            & 199                             & 75                              & 425                             & 699                                & 221.77                   \\
gpt-3.5-turbo                     & 188                             & 73                              & 429                             & 690                                & 167.77                   \\
codellama:34b-instruct            & 190                             & 52                              & 425                             & 667                                & 402.89                   \\
phind-codellama:34b-v2            & 182                             & 60                              & 399                             & 641                                & 488.27                   \\
codellama:7b-instruct             & 171                             & 72                              & 384                             & 627                                & 147.78                   \\
dolphin-mistral                   & 184                             & 76                              & 356                             & 616                                & 162.38                   \\
codebooga                         & 186                             & 56                              & 354                             & 596                                & 473.76                   \\
llama2:70b                        & 168                             & 55                              & 342                             & 565                                & 790.84                   \\
\cellcolor[HTML]{EFEFEF}\textbf{HeaderGen} & \cellcolor[HTML]{EFEFEF}\textbf{186}     & \cellcolor[HTML]{EFEFEF}\textbf{56}      & \cellcolor[HTML]{EFEFEF}\textbf{321}     & \cellcolor[HTML]{EFEFEF}\textbf{563}        & \cellcolor[HTML]{EFEFEF}\textbf{18.25 }                   \\
wizardcoder:13b-python            & 170                             & 74                              & 317                             & 561                                & 234.14                   \\
llama2:13b                        & 153                             & 40                              & 283                             & 476                                & 266.59                   \\
mistral:instruct                  & 155                             & 45                              & 250                             & 450                                & 203.78                   \\
mistral:v0.2                      & 155                             & 45                              & 248                             & 448                                & 204.60                    \\
vicuna:13b                        & 153                             & 35                              & 260                             & 448                                & 252.45                   \\
vicuna:33b                        & 133                             & 29                              & 267                             & 429                                & 434.82                   \\
wizardcoder:7b-python             & 103                             & 48                              & 254                             & 405                                & 156.62                   \\
llama2:7b                         & 140                             & 34                              & 216                             & 390                                & 146.14                   \\
\cellcolor[HTML]{EFEFEF}\textbf{HiTyper} & \cellcolor[HTML]{EFEFEF}\textbf{163}     & \cellcolor[HTML]{EFEFEF}\textbf{27}      & \cellcolor[HTML]{EFEFEF}\textbf{179}     & \cellcolor[HTML]{EFEFEF}\textbf{369}        & \cellcolor[HTML]{EFEFEF}\textbf{268.40}                   \\
wizardcoder:34b-python            & 140                             & 43                              & 178                             & 361                                & 463.05                   \\
orca2:7b                          & 117                             & 27                              & 184                             & 328                                & 215.53                   \\
vicuna:7b                         & 131                             & 17                              & 172                             & 320                                & 154.28                   \\
orca2:13b                         & 113                             & 19                              & 166                             & 298                                & 397.66                   \\
tinyllama                         & 3                               & 0                               & 23                              & 26                                 & 232.67                   \\
phind-codellama:34b-python        & 5                               & 0                               & 15                              & 20                                 & 407.20                    \\
codellama:13b-python              & 0                               & 0                               & 0                               & 0                                  & 147.21                   \\
codellama:34b-python              & 0                               & 0                               & 0                               & 0                                  & 305.74                   \\
codellama:7b-python               & 0                               & 0                               & 0                               & 0                                  & 243.01                  

		\\
		\bottomrule                    	                                
	\end{tabular}
\vspace{-.5cm}
\end{table}

%% file: discussions.tex
\section{Discussion}
\label{sec:discussion}
Similar to findings in previous work~\cite{scopeofGPT, sun2023automatic}, we observe that the construction of callgraphs does not yet significantly benefit from the use of LLMs.
In comparison to LLMs, for this task traditional SA methods remain more efficient.
However, fine-tuning GPT models showed promising improvements in callgraph analysis results, paving the way for future research in this direction.

In the type-inference tasks, LLMs such as gpt-4 and gpt-3.5, have demonstrated promising results, as evidenced in our study involving the \typeevalpy framework.
Nonetheless, in extensive Python projects using LLMs for type inference can be resource-intensive.
Moreover, employing OpenAI's services incurs monetary costs and lacks privacy for proprietary projects.
Open-source LLMs like CodeLLaMA avoid these problems as they are free and also offer the advantage of local deployment.

The LLMs tested in this study are predominantly large, having over seven billion parameters.
This renders them unsuitable for deployment on standard machines equipped with a single GPU.
In contrast, \pycg and \headergen, both traditional SA methods, are capable of operating well within such hardware constraints.
Consequently, for SA tasks, traditional SA methods still yield the best trade-off between accuracy and speed.
Nonetheless, as indicated by our findings related to type inference, where accuracy is paramount, LLMs can be effectively used, especially with fine-tuning.

\section{Threats to Validity}
\label{sec:ttv}
We list limitations and threats to the validity of our study as follows:
(1) We only analyzed the source code of the main program, excluding the code of the imported modules in the prompt. This decision was due to the complexities of constructing a prompt that accounts for the diverse import statement variations.
This particularly affects programs in the ``\texttt{imports}'' category of the \typeevalpy, \headergen, and \pycg benchmarks.
Despite this, the affected portion is relatively small (5.6\% of the total facts), so the overall results are only insignificantly altered. For a more comprehensive analysis, future work should include imported files.
(2) We used the same prompt for all models, which may not have extracted the best possible performance from each. 
(3) Open-source models often deviate from the required output formats. We addressed this by manually identifying response patterns and adding a preprocessing step for format standardization. However, this does not cover all possibilities. This issue further highlights the LLMs' inability to produce structured data consistently.

%% file: future_work_conclusion.tex
\section{Conclusion}
\label{sec:conclusion}
In this paper, we used micro-benchmarks to evaluate the application of LLMs in static analysis tasks on Python programs.
Our findings reveal that LLMs, including OpenAI's GPT-3.5 Turbo, GPT-4, and open-source models like LLaMA and CodeLLaMA, demonstrate promising capabilities in type inference, often surpassing traditional static analyses.
GPT-4 stood out as the most effective model without fine-tuning, while fine-tuning GPT-3.5 Turbo yielded significant improvements.
However, in the area of callgraph analysis, traditional methods still outperform LLMs, indicating a need for more focused fine-tuning and task-specific model adaptation.

Notably, these advancements come with substantial computational and monetary requirements.
To reduce LLM size and enhance inference speeds, future research should explore model compression techniques, such as quantization~\cite{zhuSurveyModelCompression2023}.
Further avenues of research include applying explainability methods to understand the challenges faced by LLMs in static analysis, expanding the scope to cover various static analysis tasks and programming languages, and evaluating the performance of fine-tuned open-source models.
These efforts aim to optimize LLMs for broader utility and efficiency in various static analysis tasks.

%% file: ICSE-FORGE-TypeEvalPy-LLMs.bbl

\begin{thebibliography}{23}


\ifx \showCODEN    \undefined \def \showCODEN     #1{\unskip}     \fi
\ifx \showDOI      \undefined \def \showDOI       #1{#1}\fi
\ifx \showISBNx    \undefined \def \showISBNx     #1{\unskip}     \fi
\ifx \showISBNxiii \undefined \def \showISBNxiii  #1{\unskip}     \fi
\ifx \showISSN     \undefined \def \showISSN      #1{\unskip}     \fi
\ifx \showLCCN     \undefined \def \showLCCN      #1{\unskip}     \fi
\ifx \shownote     \undefined \def \shownote      #1{#1}          \fi
\ifx \showarticletitle \undefined \def \showarticletitle #1{#1}   \fi
\ifx \showURL      \undefined \def \showURL       {\relax}        \fi
\providecommand\bibfield[2]{#2}
\providecommand\bibinfo[2]{#2}
\providecommand\natexlab[1]{#1}
\providecommand\showeprint[2][]{arXiv:#2}

\bibitem[Hug({[n.\,d.]})]%
        {HuggingFaceAI}
 \bibinfo{year}{[n.\,d.]}\natexlab{}.
\newblock \bibinfo{title}{Hugging {{Face}} {\textendash} {{The AI}} Community
  Building the Future.}
\newblock \bibinfo{howpublished}{\url{https://huggingface.co/}}.
\newblock


\bibitem[Lan({[n.\,d.]})]%
        {LangchainaiLangchainBuilding}
 \bibinfo{year}{[n.\,d.]}\natexlab{}.
\newblock \bibinfo{title}{Langchain-Ai/Langchain: {{Building}} Applications
  with {{LLMs}} through Composability}.
\newblock
  \bibinfo{howpublished}{\url{https://github.com/langchain-ai/langchain}}.
\newblock


\bibitem[Oll({[n.\,d.]})]%
        {Ollama}
 \bibinfo{year}{[n.\,d.]}\natexlab{}.
\newblock \bibinfo{title}{Ollama}.
\newblock \bibinfo{howpublished}{\url{https://ollama.ai}}.
\newblock


\bibitem[Allamanis et~al\mbox{.}(2020)]%
        {Typilus}
\bibfield{author}{\bibinfo{person}{Miltiadis Allamanis},
  \bibinfo{person}{Earl~T. Barr}, \bibinfo{person}{Soline Ducousso}, {and}
  \bibinfo{person}{Zheng Gao}.} \bibinfo{year}{2020}\natexlab{}.
\newblock \showarticletitle{Typilus: {{Neural}} Type Hints}
  \emph{(\bibinfo{series}{{{PLDI}} 2020})}. \bibinfo{publisher}{{Association
  for Computing Machinery}}, \bibinfo{address}{{New York, NY, USA}},
  \bibinfo{pages}{91--105}.
\newblock
\showISBNx{978-1-4503-7613-6}
\urldef\tempurl%
\url{https://doi.org/10.1145/3385412.3385997}
\showDOI{\tempurl}


\bibitem[Chen et~al\mbox{.}(2023)]%
        {chenUnleashingPotentialPrompt2023}
\bibfield{author}{\bibinfo{person}{Banghao Chen}, \bibinfo{person}{Zhaofeng
  Zhang}, \bibinfo{person}{Nicolas Langren{\'e}}, {and}
  \bibinfo{person}{Shengxin Zhu}.} \bibinfo{year}{2023}\natexlab{}.
\newblock \bibinfo{title}{Unleashing the Potential of Prompt Engineering in
  {{Large Language Models}}: A Comprehensive Review}.
\newblock
\newblock
\showeprint[arxiv]{2310.14735}~[cs]


\bibitem[Devlin et~al\mbox{.}(2019)]%
        {devlinBERTPretrainingDeep2019a}
\bibfield{author}{\bibinfo{person}{Jacob Devlin}, \bibinfo{person}{Ming-Wei
  Chang}, \bibinfo{person}{Kenton Lee}, {and} \bibinfo{person}{Kristina
  Toutanova}.} \bibinfo{year}{2019}\natexlab{}.
\newblock \bibinfo{title}{{{BERT}}: {{Pre-training}} of {{Deep Bidirectional
  Transformers}} for {{Language Understanding}}}.
\newblock
\newblock
\showeprint[arxiv]{1810.04805}~[cs]


\bibitem[Fan et~al\mbox{.}(2023)]%
        {fanLargeLanguageModels2023}
\bibfield{author}{\bibinfo{person}{Angela Fan}, \bibinfo{person}{Beliz
  Gokkaya}, \bibinfo{person}{Mark Harman}, \bibinfo{person}{Mitya Lyubarskiy},
  \bibinfo{person}{Shubho Sengupta}, \bibinfo{person}{Shin Yoo}, {and}
  \bibinfo{person}{Jie~M. Zhang}.} \bibinfo{year}{2023}\natexlab{}.
\newblock \bibinfo{title}{Large {{Language Models}} for {{Software
  Engineering}}: {{Survey}} and {{Open Problems}}}.
\newblock \bibinfo{howpublished}{https://arxiv.org/abs/2310.03533v4}.
\newblock


\bibitem[Hou et~al\mbox{.}(2023)]%
        {houLargeLanguageModels2023}
\bibfield{author}{\bibinfo{person}{Xinyi Hou}, \bibinfo{person}{Yanjie Zhao},
  \bibinfo{person}{Yue Liu}, \bibinfo{person}{Zhou Yang},
  \bibinfo{person}{Kailong Wang}, \bibinfo{person}{Li Li},
  \bibinfo{person}{Xiapu Luo}, \bibinfo{person}{David Lo},
  \bibinfo{person}{John Grundy}, {and} \bibinfo{person}{Haoyu Wang}.}
  \bibinfo{year}{2023}\natexlab{}.
\newblock \bibinfo{title}{Large {{Language Models}} for {{Software
  Engineering}}: {{A Systematic Literature Review}}}.
\newblock
\newblock
\urldef\tempurl%
\url{https://doi.org/10.48550/arXiv.2308.10620}
\showDOI{\tempurl}
\showeprint[arxiv]{2308.10620}~[cs]


\bibitem[Li et~al\mbox{.}(2023a)]%
        {10.1145/3611643.3613078}
\bibfield{author}{\bibinfo{person}{Haonan Li}, \bibinfo{person}{Yu Hao},
  \bibinfo{person}{Yizhuo Zhai}, {and} \bibinfo{person}{Zhiyun Qian}.}
  \bibinfo{year}{2023}\natexlab{a}.
\newblock \showarticletitle{Assisting Static Analysis with Large Language
  Models: A ChatGPT Experiment}. In \bibinfo{booktitle}{\emph{Proceedings of
  the 31st ACM Joint European Software Engineering Conference and Symposium on
  the Foundations of Software Engineering}} (<conf-loc>, <city>San
  Francisco</city>, <state>CA</state>, <country>USA</country>, </conf-loc>)
  \emph{(\bibinfo{series}{ESEC/FSE 2023})}. \bibinfo{publisher}{Association for
  Computing Machinery}, \bibinfo{address}{New York, NY, USA},
  \bibinfo{pages}{2107–2111}.
\newblock
\showISBNx{9798400703270}
\urldef\tempurl%
\url{https://doi.org/10.1145/3611643.3613078}
\showDOI{\tempurl}


\bibitem[Li et~al\mbox{.}(2023b)]%
        {liHitchhikerGuideProgram2023}
\bibfield{author}{\bibinfo{person}{Haonan Li}, \bibinfo{person}{Yu Hao},
  \bibinfo{person}{Yizhuo Zhai}, {and} \bibinfo{person}{Zhiyun Qian}.}
  \bibinfo{year}{2023}\natexlab{b}.
\newblock \bibinfo{title}{The {{Hitchhiker}}'s {{Guide}} to {{Program
  Analysis}}: {{A Journey}} with {{Large Language Models}}}.
\newblock
\newblock
\urldef\tempurl%
\url{https://doi.org/10.48550/arXiv.2308.00245}
\showDOI{\tempurl}
\showeprint[arxiv]{2308.00245}~[cs]


\bibitem[Ma et~al\mbox{.}(2023)]%
        {scopeofGPT}
\bibfield{author}{\bibinfo{person}{Wei Ma}, \bibinfo{person}{Shangqing Liu},
  \bibinfo{person}{Wang Wenhan}, \bibinfo{person}{Qiang Hu},
  \bibinfo{person}{Ye Liu}, \bibinfo{person}{Cen Zhang},
  \bibinfo{person}{Liming Nie}, {and} \bibinfo{person}{Yang Liu}.}
  \bibinfo{year}{2023}\natexlab{}.
\newblock \bibinfo{title}{The Scope of ChatGPT in Software Engineering: A
  Thorough Investigation}.
\newblock
\newblock


\bibitem[Mir et~al\mbox{.}(2022)]%
        {mirType4PyPracticalDeep2022c}
\bibfield{author}{\bibinfo{person}{Amir~M. Mir}, \bibinfo{person}{Evaldas
  Lato{\v s}kinas}, \bibinfo{person}{Sebastian Proksch}, {and}
  \bibinfo{person}{Georgios Gousios}.} \bibinfo{year}{2022}\natexlab{}.
\newblock \showarticletitle{{{Type4Py}}: Practical Deep Similarity
  Learning-Based Type Inference for Python}. In
  \bibinfo{booktitle}{\emph{Proceedings of the 44th {{International
  Conference}} on {{Software Engineering}}}} \emph{(\bibinfo{series}{{{ICSE}}
  '22})}. \bibinfo{publisher}{{Association for Computing Machinery}},
  \bibinfo{address}{{New York, NY, USA}}, \bibinfo{pages}{2241--2252}.
\newblock
\showISBNx{978-1-4503-9221-1}
\urldef\tempurl%
\url{https://doi.org/10.1145/3510003.3510124}
\showDOI{\tempurl}


\bibitem[Peng et~al\mbox{.}(2022)]%
        {HiTyper}
\bibfield{author}{\bibinfo{person}{Yun Peng}, \bibinfo{person}{Cuiyun Gao},
  \bibinfo{person}{Zongjie Li}, \bibinfo{person}{Bowei Gao},
  \bibinfo{person}{David Lo}, \bibinfo{person}{Qirun Zhang}, {and}
  \bibinfo{person}{Michael Lyu}.} \bibinfo{year}{2022}\natexlab{}.
\newblock \showarticletitle{Static Inference Meets Deep Learning: {{A}} Hybrid
  Type Inference Approach for Python}. In \bibinfo{booktitle}{\emph{Proceedings
  of the 44th International Conference on Software Engineering}}
  \emph{(\bibinfo{series}{{{ICSE}} '22})}. \bibinfo{publisher}{{Association for
  Computing Machinery}}, \bibinfo{address}{{New York, NY, USA}},
  \bibinfo{pages}{2019--2030}.
\newblock
\showISBNx{978-1-4503-9221-1}
\urldef\tempurl%
\url{https://doi.org/10.1145/3510003.3510038}
\showDOI{\tempurl}


\bibitem[Radford et~al\mbox{.}({[n.\,d.]})]%
        {radfordLanguageModelsAre}
\bibfield{author}{\bibinfo{person}{Alec Radford}, \bibinfo{person}{Jeffrey Wu},
  \bibinfo{person}{Rewon Child}, \bibinfo{person}{David Luan},
  \bibinfo{person}{Dario Amodei}, {and} \bibinfo{person}{Ilya Sutskever}.}
  \bibinfo{year}{[n.\,d.]}\natexlab{}.
\newblock \showarticletitle{Language {{Models}} Are {{Unsupervised Multitask
  Learners}}}.
\newblock  (\bibinfo{year}{[n.\,d.]}).
\newblock


\bibitem[Raffel et~al\mbox{.}(2023)]%
        {raffelExploringLimitsTransfer2023}
\bibfield{author}{\bibinfo{person}{Colin Raffel}, \bibinfo{person}{Noam
  Shazeer}, \bibinfo{person}{Adam Roberts}, \bibinfo{person}{Katherine Lee},
  \bibinfo{person}{Sharan Narang}, \bibinfo{person}{Michael Matena},
  \bibinfo{person}{Yanqi Zhou}, \bibinfo{person}{Wei Li}, {and}
  \bibinfo{person}{Peter~J. Liu}.} \bibinfo{year}{2023}\natexlab{}.
\newblock \bibinfo{title}{Exploring the {{Limits}} of {{Transfer Learning}}
  with a {{Unified Text-to-Text Transformer}}}.
\newblock
\newblock
\urldef\tempurl%
\url{https://doi.org/10.48550/arXiv.1910.10683}
\showDOI{\tempurl}
\showeprint[arxiv]{1910.10683}~[cs, stat]


\bibitem[Salis et~al\mbox{.}(2021)]%
        {salisPyCGPracticalCall2021c}
\bibfield{author}{\bibinfo{person}{Vitalis Salis}, \bibinfo{person}{Thodoris
  Sotiropoulos}, \bibinfo{person}{Panos Louridas}, \bibinfo{person}{Diomidis
  Spinellis}, {and} \bibinfo{person}{Dimitris Mitropoulos}.}
  \bibinfo{year}{2021}\natexlab{}.
\newblock \showarticletitle{{{PyCG}}: {{Practical Call Graph Generation}} in
  {{Python}}}. In \bibinfo{booktitle}{\emph{2021 {{IEEE}}/{{ACM}} 43rd
  {{International Conference}} on {{Software Engineering}} ({{ICSE}})}}.
  \bibinfo{pages}{1646--1657}.
\newblock
\showISSN{1558-1225}
\urldef\tempurl%
\url{https://doi.org/10.1109/ICSE43902.2021.00146}
\showDOI{\tempurl}


\bibitem[Seidel et~al\mbox{.}(2023)]%
        {seidel2023learning}
\bibfield{author}{\bibinfo{person}{Lukas Seidel}, \bibinfo{person}{Sedick
  David~Baker Effendi}, \bibinfo{person}{Xavier Pinho}, \bibinfo{person}{Konrad
  Rieck}, \bibinfo{person}{Brink van~der Merwe}, {and} \bibinfo{person}{Fabian
  Yamaguchi}.} \bibinfo{year}{2023}\natexlab{}.
\newblock \bibinfo{title}{Learning Type Inference for Enhanced Dataflow
  Analysis}.
\newblock
\newblock
\showeprint[arxiv]{2310.00673}~[cs.LG]


\bibitem[Sun et~al\mbox{.}(2023)]%
        {sun2023automatic}
\bibfield{author}{\bibinfo{person}{Weisong Sun}, \bibinfo{person}{Chunrong
  Fang}, \bibinfo{person}{Yudu You}, \bibinfo{person}{Yun Miao},
  \bibinfo{person}{Yi Liu}, \bibinfo{person}{Yuekang Li},
  \bibinfo{person}{Gelei Deng}, \bibinfo{person}{Shenghan Huang},
  \bibinfo{person}{Yuchen Chen}, \bibinfo{person}{Quanjun Zhang},
  \bibinfo{person}{Hanwei Qian}, \bibinfo{person}{Yang Liu}, {and}
  \bibinfo{person}{Zhenyu Chen}.} \bibinfo{year}{2023}\natexlab{}.
\newblock \bibinfo{title}{Automatic Code Summarization via ChatGPT: How Far Are
  We?}
\newblock
\newblock
\showeprint[arxiv]{2305.12865}~[cs.SE]


\bibitem[Venkatesh et~al\mbox{.}(2023a)]%
        {venkateshTypeEvalPyMicrobenchmarkingFramework2023}
\bibfield{author}{\bibinfo{person}{Ashwin Prasad~Shivarpatna Venkatesh},
  \bibinfo{person}{Samkutty Sabu}, \bibinfo{person}{Jiawei Wang},
  \bibinfo{person}{Amir~M. Mir}, \bibinfo{person}{Li Li}, {and}
  \bibinfo{person}{Eric Bodden}.} \bibinfo{year}{2023}\natexlab{a}.
\newblock \bibinfo{title}{{{TypeEvalPy}}: {{A Micro-benchmarking Framework}}
  for {{Python Type Inference Tools}}}.
\newblock
\newblock
\urldef\tempurl%
\url{https://doi.org/10.48550/arXiv.2312.16882}
\showDOI{\tempurl}
\showeprint[arxiv]{2312.16882}~[cs]


\bibitem[Venkatesh et~al\mbox{.}(2023b)]%
        {venkateshEnhancingComprehensionNavigation2023a}
\bibfield{author}{\bibinfo{person}{Ashwin Prasad~Shivarpatna Venkatesh},
  \bibinfo{person}{Jiawei Wang}, \bibinfo{person}{Li Li}, {and}
  \bibinfo{person}{Eric Bodden}.} \bibinfo{year}{2023}\natexlab{b}.
\newblock \showarticletitle{Enhancing {{Comprehension}} and {{Navigation}} in
  {{Jupyter Notebooks}} with {{Static Analysis}}}. In
  \bibinfo{booktitle}{\emph{2023 {{IEEE International Conference}} on
  {{Software Analysis}}, {{Evolution}} and {{Reengineering}} ({{SANER}})}}.
  \bibinfo{publisher}{{IEEE Computer Society}}, \bibinfo{pages}{391--401}.
\newblock
\showISBNx{978-1-66545-278-6}
\urldef\tempurl%
\url{https://doi.org/10.1109/SANER56733.2023.00044}
\showDOI{\tempurl}


\bibitem[Zhang et~al\mbox{.}(2023)]%
        {zhangSurveyLargeLanguage2023}
\bibfield{author}{\bibinfo{person}{Quanjun Zhang}, \bibinfo{person}{Chunrong
  Fang}, \bibinfo{person}{Yang Xie}, \bibinfo{person}{Yaxin Zhang},
  \bibinfo{person}{Yun Yang}, \bibinfo{person}{Weisong Sun},
  \bibinfo{person}{Shengcheng Yu}, {and} \bibinfo{person}{Zhenyu Chen}.}
  \bibinfo{year}{2023}\natexlab{}.
\newblock \bibinfo{title}{A {{Survey}} on {{Large Language Models}} for
  {{Software Engineering}}}.
\newblock
\newblock
\urldef\tempurl%
\url{https://doi.org/10.48550/arXiv.2312.15223}
\showDOI{\tempurl}
\showeprint[arxiv]{2312.15223}~[cs]


\bibitem[Zheng et~al\mbox{.}(2023)]%
        {zhengUnderstandingLargeLanguage2023}
\bibfield{author}{\bibinfo{person}{Zibin Zheng}, \bibinfo{person}{Kaiwen Ning},
  \bibinfo{person}{Jiachi Chen}, \bibinfo{person}{Yanlin Wang},
  \bibinfo{person}{Wenqing Chen}, \bibinfo{person}{Lianghong Guo}, {and}
  \bibinfo{person}{Weicheng Wang}.} \bibinfo{year}{2023}\natexlab{}.
\newblock \bibinfo{title}{Towards an {{Understanding}} of {{Large Language
  Models}} in {{Software Engineering Tasks}}}.
\newblock
\newblock
\urldef\tempurl%
\url{https://doi.org/10.48550/arXiv.2308.11396}
\showDOI{\tempurl}
\showeprint[arxiv]{2308.11396}~[cs]


\bibitem[Zhu et~al\mbox{.}(2023)]%
        {zhuSurveyModelCompression2023}
\bibfield{author}{\bibinfo{person}{Xunyu Zhu}, \bibinfo{person}{Jian Li},
  \bibinfo{person}{Yong Liu}, \bibinfo{person}{Can Ma}, {and}
  \bibinfo{person}{Weiping Wang}.} \bibinfo{year}{2023}\natexlab{}.
\newblock \bibinfo{title}{A {{Survey}} on {{Model Compression}} for {{Large
  Language Models}}}.
\newblock
\newblock
\urldef\tempurl%
\url{https://doi.org/10.48550/arXiv.2308.07633}
\showDOI{\tempurl}
\showeprint[arxiv]{2308.07633}~[cs]


\end{thebibliography}
